\documentclass[conference]{IEEEtran}
\IEEEoverridecommandlockouts
\usepackage{cite}
\usepackage{amsmath,amssymb,amsfonts}
\usepackage{algorithmic}
\usepackage{graphicx}
\usepackage{textcomp}
\usepackage{xcolor}
\def\BibTeX{{\rm B\kern-.05em{\sc i\kern-.025em b}\kern-.08em
    T\kern-.1667em\lower.7ex\hbox{E}\kern-.125emX}}
\begin{document}

\title{Computational Astrocyence: Astrocytes encode inhibitory activity into the frequency and spatial extent of their calcium elevations\\
\thanks{This work is partially funded by the Rutgers Brain Health Institute (FP00014150). Ioannis E. Polykretis is partially funded by the Onassis Foundation Scholarship.}
}

\author{\IEEEauthorblockN{Ioannis E. Polykretis}
\IEEEauthorblockA{\textit{Computational Brain Lab} \\
\textit{Rutgers University}\\
Piscataway, NJ, USA}
\and
\IEEEauthorblockN{Vladimir A. Ivanov}
\IEEEauthorblockA{\textit{Computational Brain Lab} \\
\textit{Rutgers University}\\
Piscataway, NJ, USA}
\and
\IEEEauthorblockN{Konstantinos P. Michmizos}
\IEEEauthorblockA{\textit{Computational Brain Lab} \\
\textit{Rutgers University}\\
Piscataway, NJ, USA \\
konstantinos.michmizos@cs.rutgers.edu}

}

\maketitle

\begin{abstract}
Deciphering the complex interactions between neurotransmission and astrocytic $Ca^{2+}$ elevations is a target promising a comprehensive understanding of brain function. While the astrocytic response to excitatory synaptic activity has been extensively studied, how inhibitory activity results to intracellular $Ca^{2+}$ waves remains elusive. In this study, we developed a compartmental astrocytic model that exhibits distinct levels of responsiveness to inhibitory activity. Our model suggested that the astrocytic coverage of inhibitory terminals defines the spatial and temporal scale of their $Ca^{2+}$ elevations. Understanding the interplay between the synaptic pathways and the astrocytic responses will help us identify how astrocytes work independently and cooperatively with neurons, in health and disease.
\end{abstract}


\section{Introduction}
Breaking up the neuronal monopoly of studying the brain, the implication of glia in learning and cognition is now well established \cite{Fields}. While the contribution of astrocytes, the most abundant type of glial cells, to synaptic modulation of excitatory connections has been extensively studied \cite{Araque}, their association with inhibitory synaptic transmission still remains elusive. Inhibition is crucial for the function of neural networks, as it has been involved in generating feedforward and feedback loops \cite{Buszaki} and preventing hyper-excitability \cite{Cui}. Interestingly, astrocytes are now known to detect firing activity of inhibitory neurons, which release gamma aminobutyric acid (GABA), and respond to it through intracellular $Ca^{2+}$ elevations of distinct temporal (repetitive short or prolonged) and spatial (highly localized or expanded) scales \cite{Kang, Meier, Nillson}. These $Ca^{2+}$ waves have recently been strongly related to mental dysfunctions \cite{Tian, Wetherington, Verkhratsky, Zoppo}. In fact, most brain diseases are now considered as conditions where both neurons and astrocytes are affected \cite{Barres}. In support of this notion, numerous animal model studies \cite{Seifert, Jabs, Oberheim, Jo} report alterations of astrocytic activity in a diseased brain state. However, the exact astrocytic mechanisms for relating the inhibitory synaptic transmission to the temporal and spatial characteristics of   $Ca^{2+}$ elevations are not known.

In our attempt to decipher the role of inhibitory synaptic transmission in healthy and impaired astrocytic functions, we modified our recent biologically plausible compartmental model of an astrocyte \cite{ICONS} to include $GABA_{B}$ receptors, in alignment with experimental results \cite{Kang, Meier}. Our modeling results suggested that astrocytes exhibit intensified responsiveness to short trains of pulses, increasing the frequency of their $Ca^{2+}$ elevations. Our results suggested that the extent of the astrocytic coverage of inhibitory axonal terminals is a decisive factor for the distinct spatial and temporal extent of astrocytic $Ca^{2+}$ waves. Specifically, limited coverage of active inhibitory terminals induced short and repetitive $Ca^{2+}$ elevations, which remain spatially constrained. As the coverage of terminals increased, an isolated $Ca^{2+}$ increase superimposed to $Ca^{2+}$ elevations from neighboring activity and triggered less frequent, but spatially more expanded $Ca^{2+}$ elevations.

\begin{figure}[t!]
\centerline{\includegraphics[width=\columnwidth]{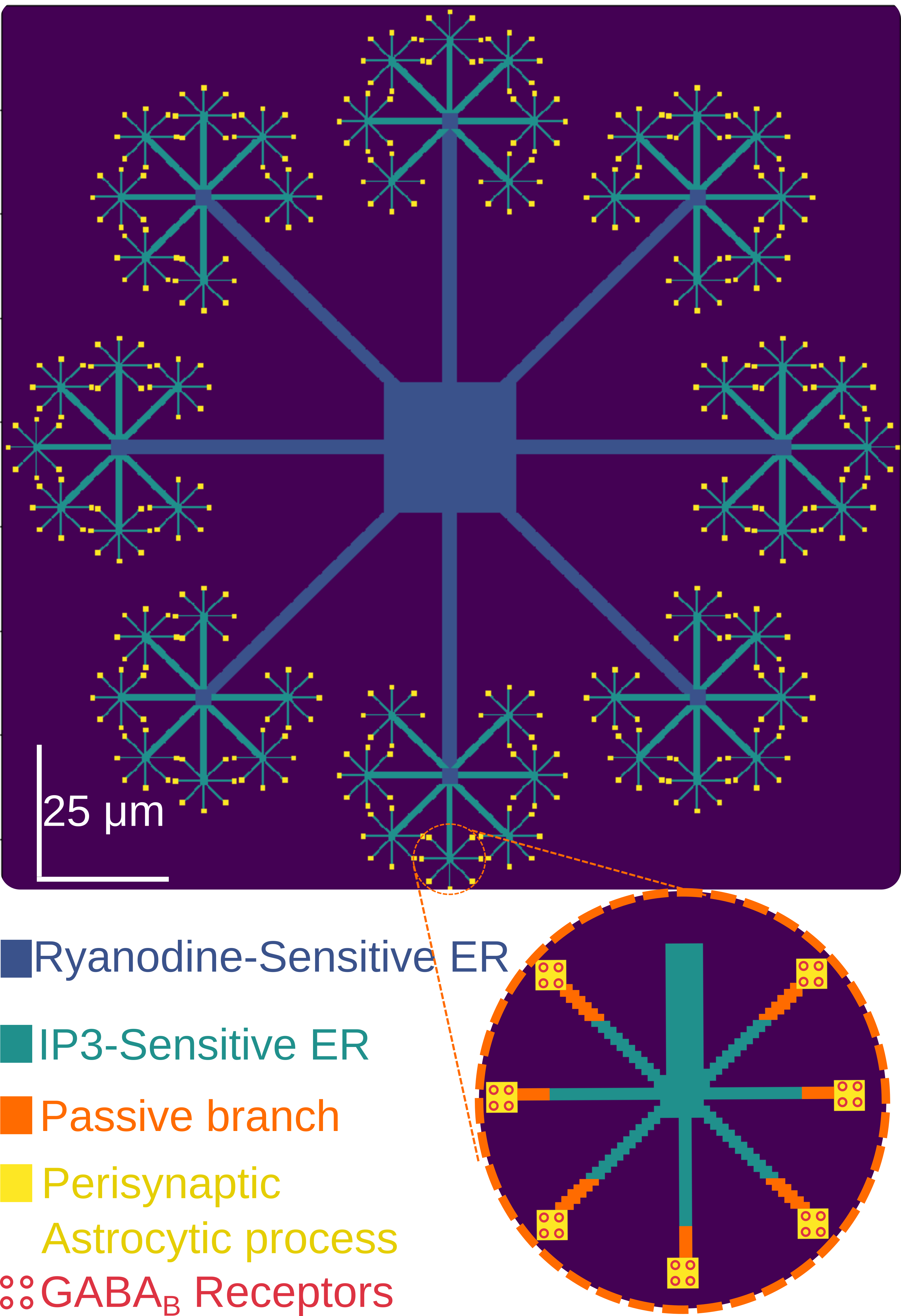}}
\caption{Geometry, compartments and their respective organelles in our 2-dimensional model of an astrocyte. $GABA_{B}$ receptors are added to our previous model \cite{ICONS}.}
\label{fig}
\end{figure}

\section{Methods}
We modified our 2-dimensional astrocytic model replicating the stellate structure of astrocytes and emulating the spatial allocation of the subcellular organelles in different compartments \cite{BI, ICONS}. Briefly, our model comprised of a somatic compartment, which extended to a number of thick branches (Fig. 1). These two compartments included both types of Endoplasmatic Reticulum (ER) reported in astrocytes; one store had receptors sensitive to Inositol Triphosphate ($IP_{3}$) and the other to Ryanodine \cite{Golovina}.  The thick branches ramified into thinner ones, which had only $IP_{3}$-sensitive ER \cite{Straub} and finally gave rise to thin tendrils, void of organelles, the perisynaptic astrocytic processes (PAPs). The assemblies of several PAPs form the now known as astrocytic microdomains (Fig. 1, inset). In the PAPs astrocytes contacted neuronal synapses, sensed the neurotransmitter spillover outside the synaptic cleft through their receptors and triggered intracellular procedures. 

The neurotransmitter receptors of astrocytes in this study are the GABA receptors (GABARs). Astrocytes have both metabotropic ($GABA_{B}Rs$) and ionotropic ($GABA_{A}Rs$) receptors \cite{Losi}. Although there is some controversy about which type of receptors gives rise to $Ca^{2+}$ elevations in astrocytes \cite{DePitta}, an increasing amount of evidence suggests that the slower and spatially expanded $Ca^{2+}$ waves are induced by $GABA_{B}Rs$ \cite{Meier}. That is why we focused on this type of G-protein-coupled receptors. This coupling of $GABA_{B}Rs$ triggers the intracellular enzymatic production of $IP_{3}$ \cite{Meier}, which then induces $Ca^{2+}$ release from the intracellular stores. We used the mathematical model initially presented in \cite{Goldberg} and further modified in \cite{ICONS} to simulate this pathway. We used a parameter $v_{GABA}$ to model the rate of agonist-dependent $IP_{3}$ production due to GABA. The numerical value that most accurately replicates the experimental findings \cite{Kang} is 6.7 $\mu$M/s. 

We used the Leaky Integrate and Fire (LIF) neuronal model to simulate the spiking activity of the interneurons. We used a membrane resistance R = 600M$\Omega$ and a time constant $\tau$ = 60ms. We adjusted the stimulation current to reproduce a widely accepted stimulation pattern that has been found to induce $Ca^{2+}$ elevations in astrocytes \cite{Kang}. In this pattern the interneurons are driven to fire a constant total number of spikes grouped in 200, 100, 66 and 33 trains of 1, 2, 3, and 6 spikes, respectively. The inter-train interval is kept constant at 2s. Specifically, we used 0.5, 0.7, 1 and 1.7 nA to drive the interneurons and force them to fire the trains of 1, 2, 3 and 6 spikes respectively.

To translate the spiking activity of the interneurons to the amount of neurotransmitter sensed by the astrocytes, we modified the model presented in \cite{Tsodyks}. Given a spiking pattern, the model predicts the amount of the neurotransmitter released inside the synaptic cleft. However, the PAP senses only a percentage of this amount that spills out of the cleft. To account for that spillover, we used only a fraction of that amount of neurotransmitter that the model predicts. This amount of stimulant was subsequently fed into the astrocytic model to trigger $IP_{3}$ production and, consequently, $Ca^{2+}$ release from the ER that can be seen as $Ca^{2+}$ elevations.

The 33-6 protocol \cite{Kang} was used for simulation brevity in all experiments, where the reproduction of another protocol was not imperative.

\section{Results}
Our astrocytic model exhibited the GABA-induced $Ca^{2+}$ waves that have been reported in astrocytes \cite{Kang, Nillson}. We observed a wide range of $Ca^{2+}$ waves emerging through the intracellular pathway described above.

\begin{figure*}[ht!]
\centerline{\includegraphics[width=\textwidth]{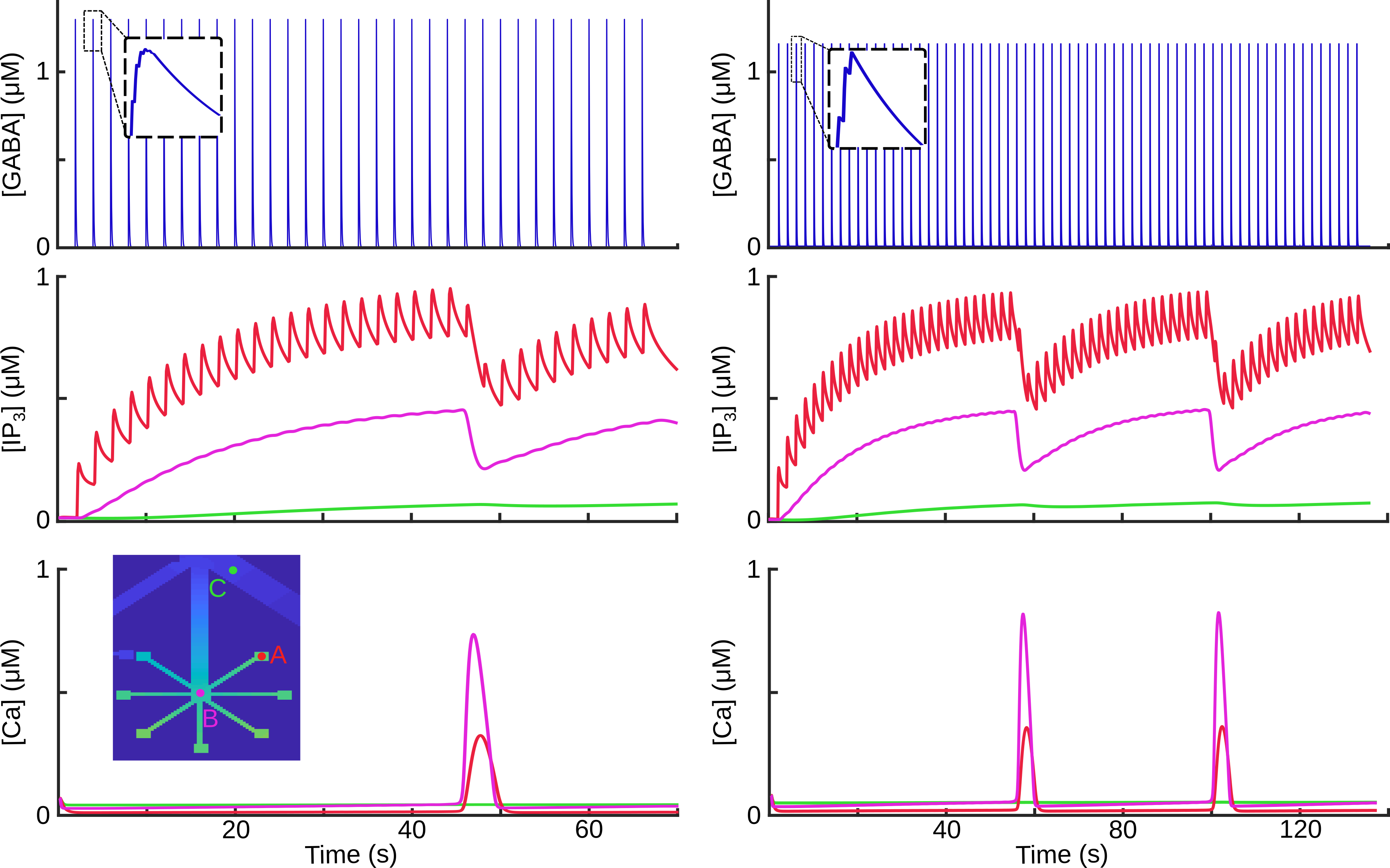}}
\caption{Astrocytic activity in response to two stimulation patterns. The GABA stimulation patterns (top) induce IP3 production (middle), which triggers Ca2+ release (bottom). Maximal response in terms of number of Ca2+ elevations and their frequency is achieved when the spikes are grouped in 66 trains of 3 (right column) and not in 33 trains of 6 (left column).}
\label{fig}
\end{figure*}

First, we examined the responsiveness of the cell to different types of stimulus . We used different stimulation protocols to drive our model and the results are shown in Figure 2. We stimulated all the PAPs of a single microdomain, assuming that they all contact the axonal terminals of a single spiking interneuron or multiple distinct interneurons driven simultaneously (Fig.2, left, bottom inset). Our results suggested that the 66-3 protocol (Fig.2, right, bottom) produced a more prominent response than the respective response to the 33-6 protocol (Fig.2, left, bottom). The 66-3 protocol gave rise to multiple $Ca^{2+}$ elevations, while the 33-6 protocol gave rise to a single elevation during the stimulation period. As seen in Fig.2 (top, insets), every spike of a single train released an amount of the available neurotransmitters. When the trains included more than three spikes, the amount of released neurotransmitter after the third spike was inconsequential for astrocytes, resulting in weaker $Ca^{2+}$ responses. We also tested the 200-1 and 100-2 stimulation protocols and the response of the cell was always less prominent (results not shown). Therefore, our model followed closely the experimental findings that show maximal astrocytic responsiveness to the 66-3 protocol (\cite{Kang}, Fig.6, c).

Elevations of different spatial extent and temporal duration in neighboring cells are described (\cite{Kang}, Fig.6, b). We simultaneously stimulated multiple microdomains and examined whether such elevations can be observed. As seen in Fig.3, stimulation of multiple inhibitory terminals induced $Ca^{2+}$ elevations that were temporally extended and spatially expanded (Fig.3 bottom, green curve).

\section{Discussion}
Our compartmental astrocytic model suggested possible relationships in space and time between the inhibitory synaptic transmission and intracellular $Ca^{2+}$ elevations. Specifically, the level and the frequency of astrocytic $Ca^{2+}$ activity depended on the stimulation pattern of the neurons. Moreover, the spatial extent and temporal duration of $Ca^{2+}$ waves depended on the number of inhibitory terminals that an astrocyte covered. 

First, our cell model aligned with experimental results \cite{Kang}, as it exhibited intensified responsiveness to short trains of inhibitory spikes by increasing the frequency of its $Ca^{2+}$ elevations. Sparse distribution of the spikes or grouping in longer trains did not induce significant $Ca^{2+}$ elevations. We propose that this strengthened response was due to the maximal exploitation of synaptic resources by the short spike trains. Conversely, sparse spikes could not fully activate the synaptic resources and longer trains depleted them, inducing weaker responses. A limited number of active synapses induced short and repetitive, spatially compartmentalized $Ca^{2+}$ elevations. The spatial extension of these elevations was confined to the stimulated microdomain (Fig.2, bottom, red and purple curves), since the thicker branches were not affected by the wave (Fig.2, bottom, green curve). However, the spatial resolution of the experimental study was not high enough, so that we could test this observation (\cite{Kang}, Fig.6, a).

\begin{figure}[ht!]
\centerline{\includegraphics[width=\columnwidth]{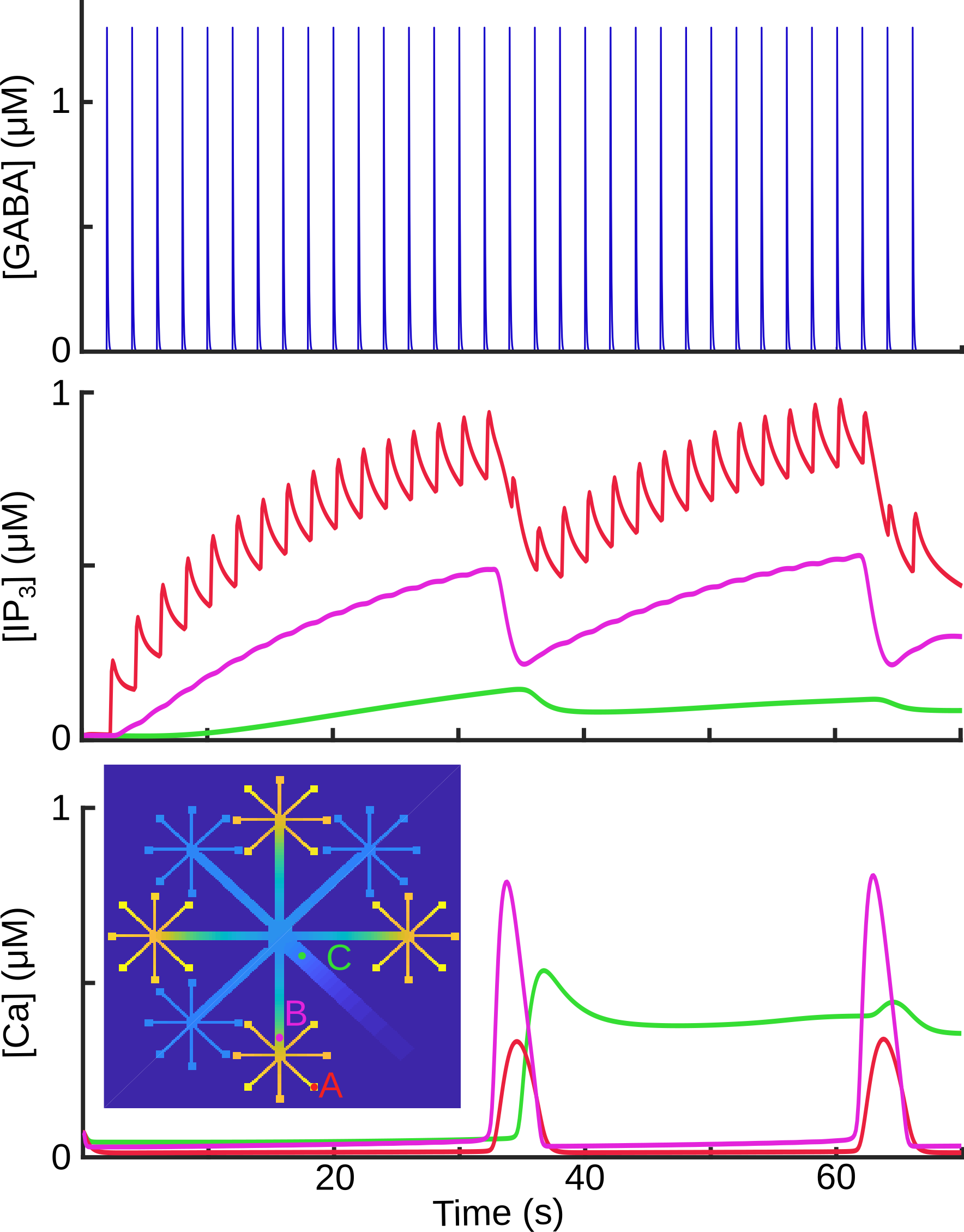}}
\caption{Prolonged and spatially expanded Ca2+ elevations observed with extensive astrocytic coverage of synapses. Astrocytic coverage of multiple spiking inhibitory terminals induces temporally extended and spatially expanded Ca2+ elevations (bottom, green).}
\label{fig}
\end{figure}

Our results exhibited spatially expanded and temporally extended $Ca^{2+}$ elevations, similar to the experimental findings (\cite{Kang}, Fig.6, b). Our model suggested that the number of inhibitory terminals that an astrocyte contacts is a decisive factor for the spatial extent and the temporal duration of astrocytic $Ca^{2+}$ elevations. Limited coverage of terminals induced brief, repetitive and localized $Ca^{2+}$ elevations. However, increasing the number of active synapses that an astrocyte ensheathes, induced simultaneous, isolated $Ca^{2+}$ elevations in multiple microdomains. Their superposition was sufficient to trigger the Ryanodine-sensitive ER in the thicker branches and give rise to elevations that were prolonged in time, spatially expanded and had lower frequency (Fig.3 bottom, green curve). 

In our previous work, we showed that astrocytes, when stimulated by excitatory neurotransmission, can induce neuronal synchronization \cite{ICONS} and spatial clustering of synaptic plasticity \cite{BI}. This study complements our previous results by showing that inhibitory neurotransmission may also shape $Ca^{2+}$ elevations in astrocytes. Taken together, our approaches may help further our understanding on the emergence of astrocytic $Ca^{2+}$ responses to the most important synaptic pathways, as well as their interaction. 

Overall, our computational work aims to study how astrocytes may work independently and cooperatively with neurons in health and disease.






\end{document}